\documentclass[]{aa}
\usepackage{graphicx}
\usepackage{txfonts}
\begin{document}
   \title{The distribution of ND$_2$H in LDN1689N}

   \subtitle{}

   \author{M. Gerin
          \inst{1}
          \and
          D.C. Lis\inst{2}
\and
S. Philipp\inst{3}
\and R. G\"usten\inst{3}
\and E. Roueff\inst{4}
\and V. Reveret\inst{5}
          }

   \offprints{M. Gerin}

   \institute{LERMA, CNRS UMR8112, Observatoire de Paris and ENS ,
              24 Rue Lhomond, 75231 Paris cedex 05 France,
              \email{gerin@lra.ens.fr}
         \and
             California Institute of Technology, MC 320-47, Pasadena,
              CA 91125 , USA,
             \email{dcl@submm.caltech.edu}
\and  
Max Planck Institut f\"ur Radioastronomie, Auf dem H\"ugel 69, Bonn, Germany,
\email{philipp,guesten@mpifr-bonn.mpg.de}
\and            
LUTH, CNRS UMR8102, Observatoire de Paris and Universit\'e Paris 7, 
Place J. Janssen, 92190 Meudon, France , 
\email{evelyne.roueff@obspm.fr}
\and
European Southern Observatory, Casilla 19001, Santiago 19, Chile, 
\email{vreveret@eso.org}
}

   \date{Received 10 April 2006 / Accepted 01 June 2006 }

 
  \abstract
  {}
  {Finding tracers of the innermost regions of prestellar cores is
important for understanding their chemical and dynamical evolution
before the onset of gravitational collapse. While classical molecular 
tracers, such as CO and CS, have been shown to be strongly depleted 
in cold, dense gas by
condensation on grain mantles, it has been a subject of discussion to
what extent nitrogen-bearing species, such as ammonia, are affected by
this process. As deuterium fractionation is efficient in cold,
dense gas, deuterated species are excellent tracers of prestellar cores. 
A comparison of the spatial distribution of neutral and ionized
deuterated species with the dust continuum emission can thus provide
important insights into the physical and chemical structure of such regions.}
  {We study the spatial distribution of the ground-state 335.5~GHz
line of  ND$_2$H in LDN1689N, using APEX, and compare it with the
distribution of the DCO$^+$(3--2) line, as well as the 
350~$\mu$m dust continuum  
emission observed with the SHARC~II bolometer camera at CSO.
}
  {While the distribution of the ND$_2$H emission in LDN1689N is
generally similar to 
that of the 350~$\mu$m  dust continuum emission, the
peak of the  ND$_2$H emission is offset by $\sim$$10''$ to the East 
from the dust continuum and DCO$^+$ emission peak. 
ND$_2$H and ND$_3$ share the same spatial
distribution. The observed offset between the ND$_2$H and DCO$^+$ emission 
is consistent with the hypothesis that the deuterium peak in LDN1689N
is an interaction region between the 
outflow shock from IRAS16293--2422 and the dense ambient gas. 
We detect the $J = 4 \rightarrow 3$ line of H$^{13}$CO$^+$ at 
346.998 GHz in the image side band serendipitously. This line 
 shows the same spatial distribution as DCO$^+$(3--2), and peaks
close to the 350~$\mu$m emission maximum  which provides further support
for the shock interaction scenario.}
  {}


   \maketitle
%

\section{Introduction}
Deuteration of nitrogen compounds, such as ammonia and N$_2$H$^+$, is 
spectacular in a number of environments including  dark clouds, such as 
L134N (Tin\'e et al. 2000; Roueff et al. 2000; Roueff et al. 2005), 
low-mass star forming 
regions and prestellar cores such as LDN1689N and Barnard~1 (Loinard et 
al. 2001; Gerin et al. 2001; Lis et al. 2002a). The discussion of the 
respective 
contributions of grain and gas-phase processes in the deuteration is 
active, but no definite solution is yet available. The lack of 
detection of deuterated water in ices toward low-mass young stellar 
objects (YSOs) by Dartois et al. (2003) and Parise et al. (2003) 
suggests that ``another mechanism than pure solid state chemistry may 
be active to produce very high deuterium enrichment in the gas phase.''

Further constraints on the deuterium fractionation mechanisms
are provided by the spatial distribution of deuterated species,
as compared with the molecular gas distribution traced
by the submillimeter dust continuum emission. Whereas maps of singly
deuterated species have been published in many cores --- 
the relatively large line intensities of e.g. DCO$^+$, N$_2$D$^+$
allow easy mapping with state of the art detectors --- mapping multiply
deuterated species has proven to be a challenge, given the relatively
low line intensities. Ceccarelli et al. (2001) have shown that the
D$_2$CO emission is extended around the class 0 protostar IRAS16293--2422.
Roueff et al. (2005) present
a limited map of the ND$_3$ ground-state transition at 309 GHz
with the CSO in LDN1689N, which unfortunately suffers from rather poor pointing
accuracy. Because the ground state ND$_2$H $1_{0,1} - 0_{0,0}$ lines,
at 335.5 for the ortho species and 335.4 GHz for the para species (See Coudert 
and Roueff (2006) for NH$_3$ and its isotopologues line frequencies), are
relatively strong (0.6 K in LDN1689N; Lis et al. 2006) and 
the atmospheric transmission is good at this frequency, we have
carried out the first map of a doubly deuterated species in a dense core. 
   
\section{Observations}
The observations have been carried out using the APEX-2a 350~GHz receiver
of the Atacama Pathfinder Experiment (APEX \footnote{This publication is based
on data acquired with the Atacama Pathfinder Experiment (APEX). APEX is a
 collaboration between the Max-Planck-Institut f\"ur Radioastronomie, the 
European Southern Observatory and the Onsala Space Observatory. }).
 We combined data obtained during two periods,
August 2005 and October 2005.
The receiver was tuned in DSB with the ND$_2$H lines in the lower sideband.
The backend was the facility MPIfR Fast Fourier Spectrometer.
We observed a total of 10 positions towards LDN1689N 
($\alpha_{2000}$ = 16:32:29.470,
$\delta_{2000}$ = --24:28:52.60). The pointing was checked regularly
and found accuracte to $\sim 2 - 3 ''$.
The observations have been taken in
the position-switched mode with a reference position located 240$''$ East
of the source for the August 2005 data. The October 2005 data have been
taken with a reference position offset by 10$'$ in azimuth. 
Line and continuum maps shown in Lis et al. (2002b) and Stark et al. (2004)
show that the source emission rapidly drops towards the East.

The data have been  corrected for the sideband gain and  
the APEX main beam efficiency of 0.7 at 335 GHz. The temperature
scale used in this paper is  the main beam brightness temperature scale.
Overall, the data calibration agree well with the CSO
spectrum presented by Lis et al. (2006). The FWHM beam size of APEX is
18$''$ at 335.5~GHz.

   \begin{figure*}
   \centering
\resizebox{9cm}{!}{
\rotatebox{270}{
   \includegraphics{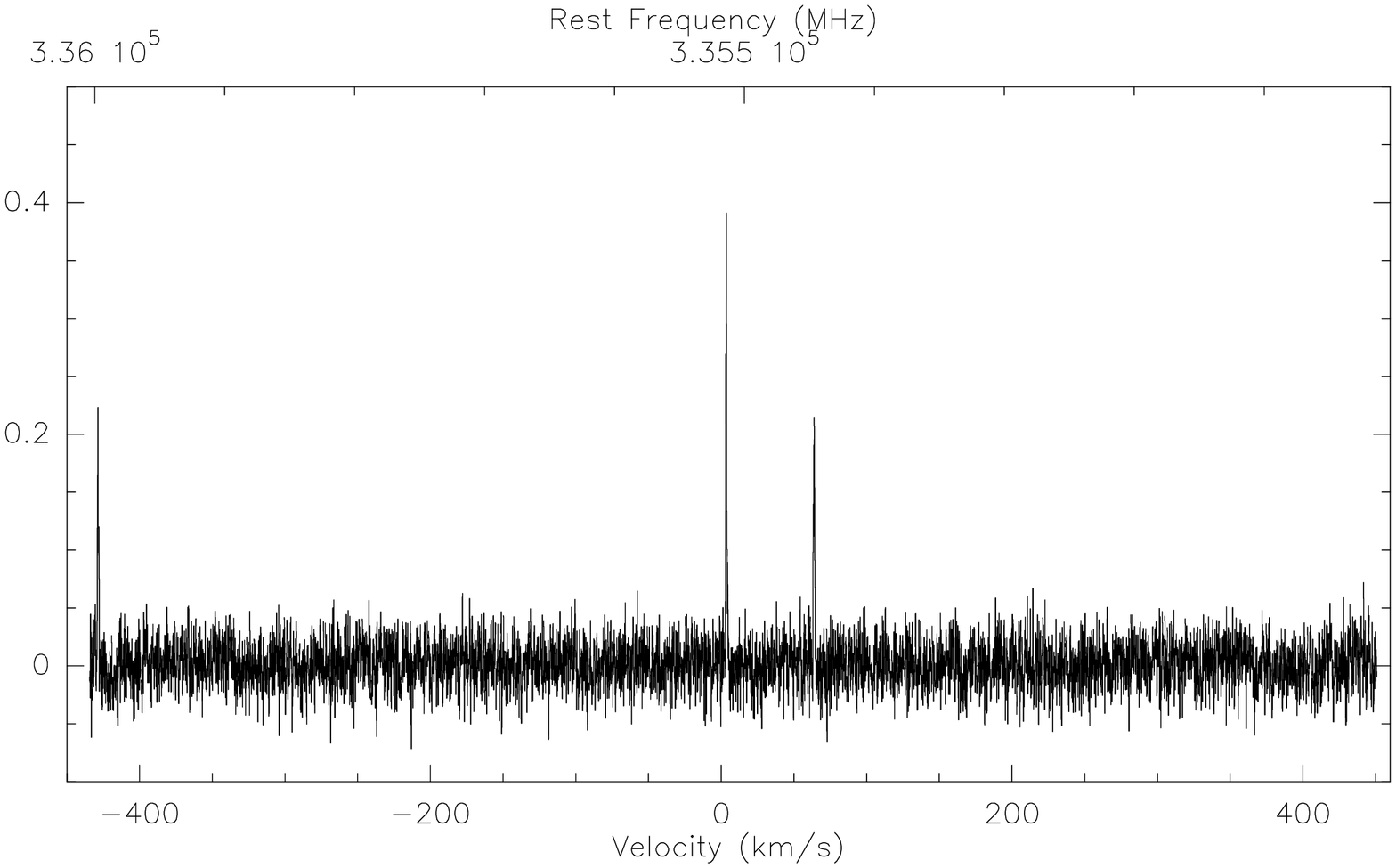}}}
\resizebox{7cm}{!}{
\rotatebox{270}{
   \includegraphics{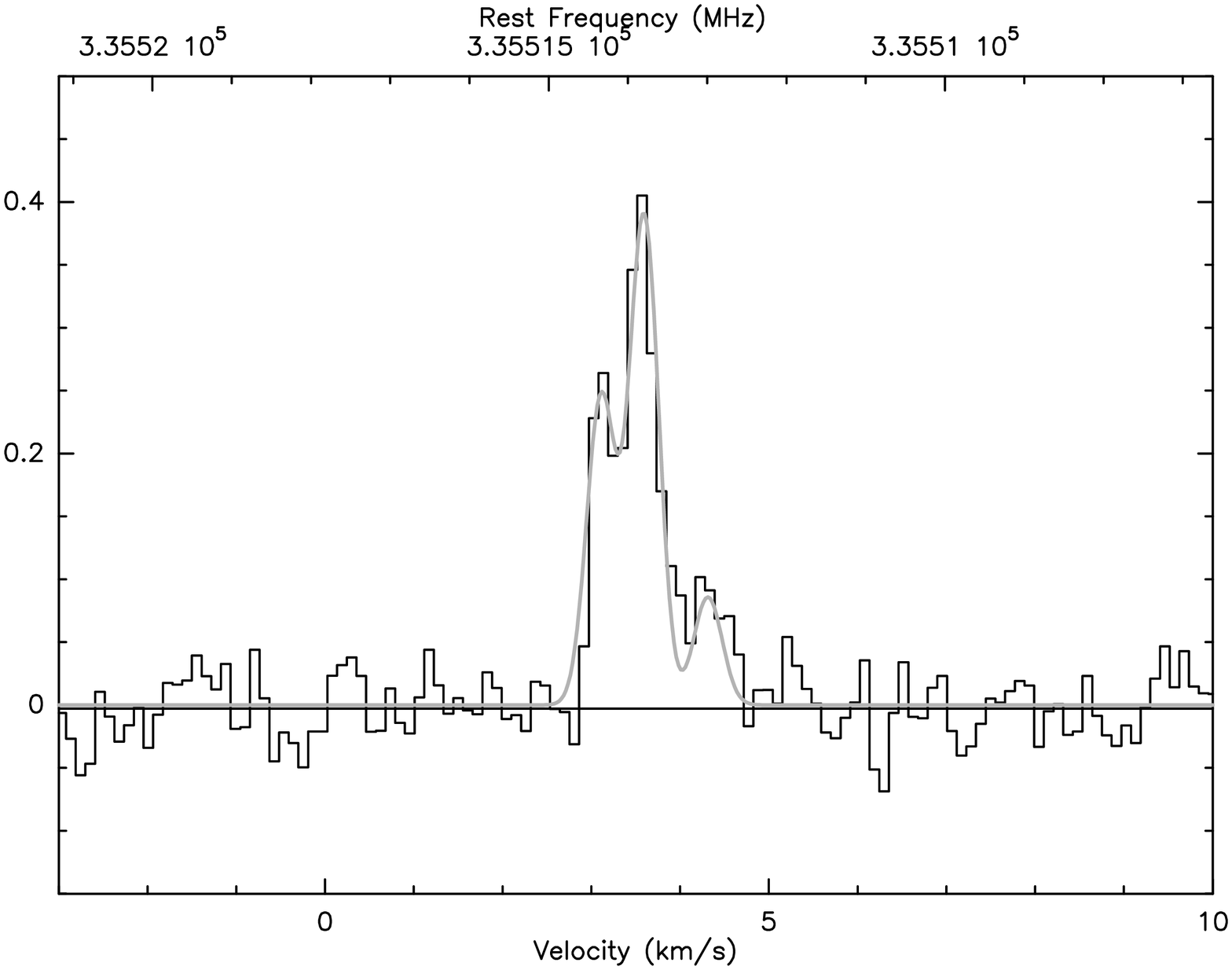}}}
   \caption{Left : Full spectrum averaged over the map. 
  p-ND$_2$H($1_{11}-0_{00}$) at 335.446 GHz, 
o-ND$_2$H$(1_{11}-0_{00}$) at 335.514 GHz and 
 H$^{13}$CO$^+$(4--3) at 346.998 GHz (in the image side band)
are detected.  
    Right : Enlargement showing a fit of the hyperfine structure
of the o-ND$_2$H line. The temperature scale is $T_{\rm mb}$ (K) and the velocity scale is relative to the LSR.}          \label{fig:spec}%
    \end{figure*}

\begin{figure*}
   \centering
\resizebox{16cm}{!}{
\rotatebox{270}{
\includegraphics{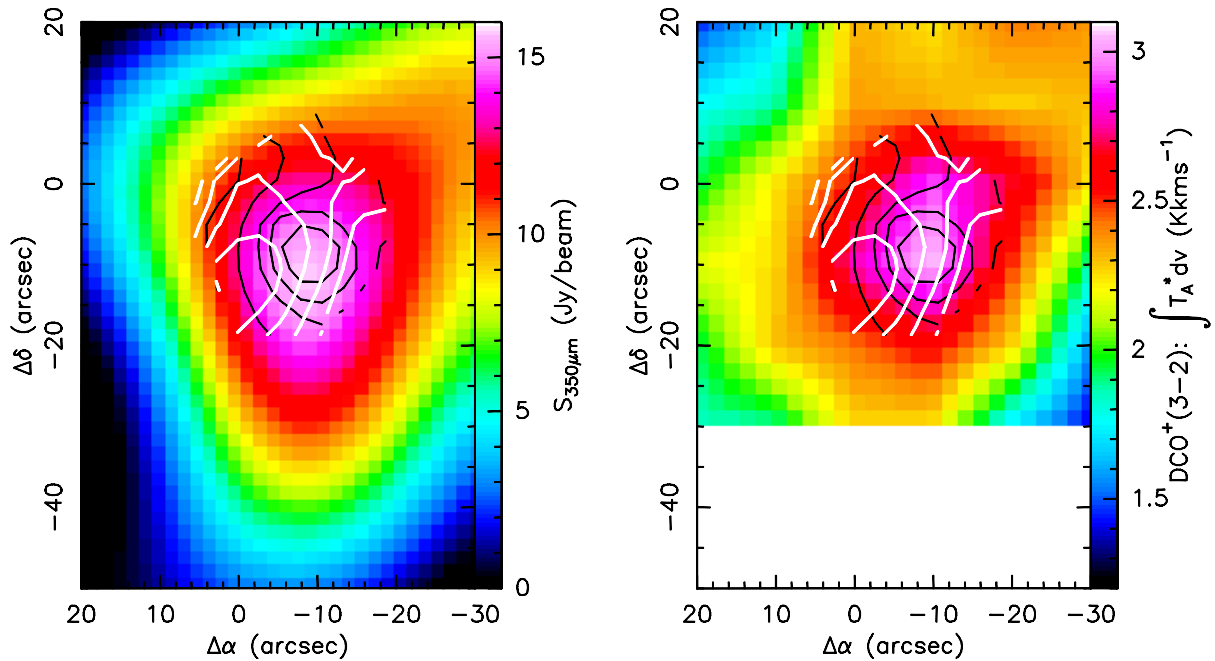}}}
   \caption{Left:
Color image of the 350~$\mu$m dust continuum emission obtained
with SHARC~II at the CSO, convolved to 20$''$ angular resolution. 
The intensity scale is given in Jy/beam.
White contours show the integrated intensity of the 
o-ND$_2$H line at 335.5~GHz, with contour levels from 0.25 to 0.45
Kkms$^{-1}$ with 0.05 Kkms$^{-1}$ spacing. Black contours 
show the integrated intensity of
the  H$^{13}$CO$^+$(4--3) line at 346.998~GHz, with  contour levels 
from 0.25 to 0.87
Kkms$^{-1}$ with 0.125 Kkms$^{-1}$ spacing. The map center is
$\alpha_{2000}$ = 16:32:29.470,  $\delta_{2000}$ = -24:28:52.60.
Right: The same contours plotted on a DCO$^+$(3--2) map (Lis et al. 2002b).
} 
              \label{fig:map}%
    \end{figure*}

\section{Results}
Fig.~\ref{fig:spec} shows
the average spectrum integrated over the map, while contour maps are shown
in Fig.~\ref{fig:map}.
The ortho- and para-ND$_2$H($1_{11}-0_{00}$) lines are clearly
detected with a mean intensity ratio of 2:1, reproducing the 
ortho-to-para statistical weight ratio (Lis et al. 2006).
We have also detected the H$^{13}$CO$^+$(4--3) line at 346.998 GHz in the
image side band, which is discussed below.

\subsection{ND$_2$H spatial distribution}
Fig.~\ref{fig:map} shows
contours of the main ND$_2$H emission at 335.5 GHz (in white) overlayed over
a 350~$\mu$m continuum image. The ND$_2$H emission
is clearly extended and shows
an elongated shape generally similar to the dust continuum emission.
The limited ND$_3$ map shown by Roueff et al. (2005) shows the
same elongated pattern.
Over the limited extend of the ND$_2$H map, the line and continuum
emissions are positively correlated with a correlation coefficient
of 0.29. This rather weak correlation is due to the
difference in peak positions of the  ND$_2$H line emission with
respect to the continuum peak, the ND$_2$H emission beeing offset
by $\sim$$10''$ to the East of the dust contimuum emission maximum.
This offset far exceeds the pointing uncertainties ($\sim$$2-3''$).

The right panel in Fig.~\ref{fig:map} shows an overlay of the ND$_2$H contours
on a DCO$^+$(3--2) map (Lis et al. 2002b). The ND$_2$H peak is clearly
offset from the DCO$^+$ peak too. D$_2$CO (Ceccarelli et al. 2000),
ND$_3$ (Roueff et al. 2005),  ND$_2$H (Loinard et al. 2001, Roueff et al. 
2005),  DCO$^+$ (Lis et al; 2002b), 
all appear at a blueshifted velocity compared to the cloud envelope : 
V$_{LSR}$ = 3.4 -- 3.6 kms$^{-1}$ versus
V$_{LSR}$ = 3.8 -- 4.0 kms$^{-1}$ for C$^{18}$O and C$^{17}$O 
(Stark et al. 2004). This velocity difference
was first recognized by Lis et al. (2002b) who suggested that it results from
the shock interaction of the powerful blue lobe of the  IRAS16293-2422 outflow
with the dense core. The blue lobe of the molecular outflow
 is interacting and compressing LDN1689N, creating  dense and cold 
post-shock gas, blueshifted relative to the cloud envelope.
 The deuterated species preferably sample this cold and 
dense material as deuterium fractionation is more efficient at low 
temperatures and high gas densities.
The large-scale CO and continuum maps presented by
Lis et al. (2002b) and Stark et al. (2004) present
convincing evidence for this scenario. 

The spatial offset between DCO$^+$ and ND$_2$H provides additional
support for the C-shock hypothesis. The DCO$^+$ ion is detected ahead of
the neutral species ND$_2$H, closer to IRAS16293-2422. 
The spatial separation, $10''$, or  $\sim$$0.01$~pc, is
consistent with C-shock models with a pre-shock density of 10$^4$ cm$^{-3}$ and
a magnetic field of 30 $\mu$G (Lesaffre et al. 2004), which predict a
total shock size of 0.04~pc. The deuterium chemistry in shocks has been 
previously studied by Pineau des For\^ets et al. (1989) and Bergin et al. 
(1999) in  different contexts.
Because of the sensitivity of deuterium fractionation to both the gas
temperature and the molecular depletion, it is expected that DCO$^+$
will be reformed more rapidly than ND$_2$H in the post-shock gas.
While DCO$^+$ is formed as soon as the gas temperature drops,
efficient ND$_2$H formation requires both cold temperatures and
significant CO depletions, which are expected to occur downstream (Bergin et
al. 1999). A spatial offset could therefore exist between abundance peaks
of DCO$^+$ and ND$_2$H, as we have detected. More detailed shock models,    
would be able to further test the validity of this scenario. 
 

\subsection{ND$_2$H abundance}
We have fitted the ND$_2$H spectra using the hyperfine fitting
 command in CLASS. Although the signal-to-noise
ratio is not high enough for securely deriving
the line opacity at all positions, it is clear that 
the maximum opacity coincides with
the maximum signal. The opacity derived for the average 
of all positions excluding (10$''$,0$''$) is $0.2 \pm 0.48$,
and the corresponding excitation temperature is 
$8 \pm 3$ K, in good agreement with Lis et al. (2006).
The spectra show some indication of higher line opacity, and lower
excitation temperature ($\tau = 3.11 \pm 1.5$; $T_{ex} = 5 \pm 1 $
towards the two positions with highest intensities (0,-10$''$) and (0,-20$''$).

The ND$_2$H column densities have been derived assuming optically thin
emission, LTE and using a uniform  excitation temperature of 5~K.
 They are listed, together with the H$_2$ column density 
derived from dust continuum
measurements in Table~\ref{tab:abund}.
We use the 350~$\mu$m  dust continuum 
map obtained with SHARC~II at the CSO (Fig.~\ref{fig:map}), and a 
a dust opacity of
$\kappa_{350} = 0.07$~cm$^{2}$g$^{-1}$ corresponding to
$\kappa_{1300} = 0.005$~cm$^{2}$g$^{-1}$
(Ossenkopf \& Henning 1994) for a dust emissivity index $\beta = 2$, 
and a dust temperature of 16~K (Stark et al. 2004). 
The 350~$\mu$m continuum fluxes are in fair agreement with 
the values reported by Stark et al. (2004), but on the lower side:
we detect a maximum intensity of 16~Jy in a 20'' beam, to be compared
with $19.4 \pm 2.3$~Jy in Stark et al. (2004). 
This dicrepancy may be an indication that some low-level entended
emission is filtered out in the SHARC~II image, which has been
obtained in the ``AC-biased mode'', without the secondary chopper. 
The
resulting H$_2$ column densities reported in Table~\ref{tab:abund}
could therefore be affected by this 35\% uncertainty.

The H$_2$ column
density ranges from $3.1 \times 10^{22}$ cm$^{-2}$ to 
$7.2 \times 10^{22}$ cm$^{-2}$ across the map. The corresponding
ND$_2$H abundances varies from $0.7 \times 10^{-9}$ to $2.6 \times 10^{-9}$
relative to H$_2$, with a mean value of $1.8 \times 10^{-9}$, in
good agreement with previous work (Lis et al. 2006).
ND$_2$H appears to be remarkably abundant in this dense core.
For the three positions where reliable ND$_3$ data
exist (Roueff et al. 2005), 
the [ND$_3$]/[ND$_2$H] abundance ratio is 0.01 -- 0.02,
and seems to increase with increasing ND$_2$H column density.

\begin{table*}
\caption{}
\label{tab:abund}
\begin{tabular}{lccccc}
Position & V & $\delta$V & N(ND$_2$H)$^1$ & N(ND$_3$)$^2$ & N(H$_2$)$^3$ \\
(arcsec) & kms$^{-1}$ & kms$^{-1}$ & 10$^{13}$ cm$^{-2}$ & 
10$^{12}$ cm$^{-2}$ & 10$^{22}$ cm$^{-2}$\\
\hline
-10,10 & $3.63 \pm 0.02$ & $0.33  \pm 0.03 $ & $9.8 \pm 2$  & &$4.1 \pm .5$\\
10,0 & $3.53 \pm 0.02$ & $0.20  \pm 0.07 $ & $3.6 \pm 1$  & &$3.1 \pm .5$\\
0,0 & $3.56 \pm 0.02$ & $0.40  \pm 0.05 $ & $12.7 \pm 2$ & 1.5 & $5.1 \pm .5$\\
-10,0 & $3.62 \pm 0.02$ & $0.39  \pm 0.05 $ & $10.0 \pm 2$  & & $5.9 \pm .5$\\
-20,0 & $3.75 \pm 0.1$ & $0.8 \pm 0.3 $ & $8.9 \pm 2$& &$5.4 \pm .5$\\
0, -10 & $3.60 \pm 0.02$ & $0.30 \pm 0.05 $ &  $15.8 \pm 2$ & & $5.8 \pm .5$\\
-10,-10 & $3.70 \pm 0.05$ & $0.55  \pm 0.1 $ & $11.7\pm 2$  & 1.8 & 
$7.1 \pm .5$\\
-20,-10 & $3.55 \pm 0.02$ & $0.26 \pm 0.05 $ & $4.6 \pm 1$& &$5.7 \pm .5$\\
0,-20 & $3.69 \pm 0.02$ & $0.45 \pm 0.06 $ & $12.7 \pm 2$& 2.6 &$5.2 \pm .5$\\
-10,-20 & $3.72 \pm 0.03$ & $0.58  \pm 0.08 $ & $8.9 \pm 2$  & &$6.5 \pm .5$\\
\hline
\end{tabular}

$^1$ Derived assuming LTE and $T_{ex} = 5$~K. $^2$ Roueff et al. (2005).$^3$ Derived assuming
$T_{dust} = 16$~K and $\kappa_{350} = 0.07$~cm$^{2}$g$^{-1}$.
\end{table*}

\subsection{H$^{13}$CO$^+$}
A relatively strong line is detected at the lower end of the velocity range
in Fig.~\ref{fig:spec}.  Careful tests have been performed using the 
CSO and APEX telescopes for identifying this line. It turns out that the
line can be assigned to the $J = 4 \rightarrow 3$ transition of H$^{13}$CO$^+$
at 346.998 GHz seen in the image side band. At APEX, when the system is
tuned to ND$_2$H, the H$^{13}$CO$^+$ 
line falls 16MHz outside the 1GHz wide IF band, 
but is aliased back into the signal band. The band edge IF 
filter attenuates the line intensity, therefore the correct
intensity scale has been established by 
comparing the original data
with additional measurements taken
with APEX on May 22nd, 2006 with the H$^{13}$CO$^+$ line centered
in the signal band. The 2005  data 
have been multiplied by 2.5 to
match the properly tuned 2006  data. We have checked that 
the spatial distributions of 
H$^{13}$CO$^+$(4-3) agree reasonably well between both datasets.    
H$^{13}$CO$^+$(4--3) has a more sharply peaked spatial distribution 
than ND$_2$H (Fig.~\ref{fig:map}), which is very similar to the DCO$^+$(3--2) 
map obtained at
the CSO (see right panel of Fig.~\ref{fig:map}).


%

\section{Conclusions}

\begin{enumerate}
\item We show that the ND$_2$H emission is extended in the \object{LDN1689N}
dense core. The abundance relative to H$_2$, derived from LTE, 
is $\sim 1.8 \times 10^{-9}$, which makes ND$_2$H a remarkably
abundant molecule in this dense core. The [ND$_3$]/[ND$_2$H] abundance ratio
is 0.01 -- 0.02.
\item The ND$_2$H emission is spatially offset from the dust continuum
DCO$^+$(3--2) and H$^{13}$CO$^+$(4--3) emission by $\sim 0.01$~pc, 
and appears to be blueshifted relative to the
cloud envelope. Both findings are qualitatively consistent with the scenario
of the formation of the LDN1689N dense core by the interaction of the
blue lobe of the  \object{IRAS16293--2422} molecular outflow with the ambient material.
Detailed models could further test this scenario.
   \end{enumerate}

\begin{acknowledgements}
Caltech Submillimeter Observatory is supported by the U.S. 
National Science Foundation, grant AST~0540882. We thank the MPG and ESO 
APEX teams for their support, especially P. Bergman and L.- \AA. Nyman.
\end{acknowledgements}

\end{document}